# Evaluation of Watson-like Integrals for Hyper bcc Antiferromagnetic Lattice

S M Radošević, M R Pantić, D V Kapor, M V Pavkov-Hrvojević and M G Škrinjar

Department of Physics, Faculty of Sciences, Trg Dositeja Obradovića 4
Novi Sad, Serbia

E-mail: `slobodan@df.uns.ac.rs`

**Abstract.** Watson-like integrals for a $d$-dimensional bcc antiferromagnetic lattice

$$I_d(\eta) = \frac{1}{\pi^d} \prod_{i=1}^{d} \int_0^{\pi} \mathrm{d}x_i \, \frac{\eta}{\sqrt{\eta^2 - \prod_{j=1}^{d}(\cos[x_j/2])^2}}$$

and

$$J_d(\eta) = \frac{1}{\pi^d} \prod_{i=1}^{d} \int_0^{\pi} \mathrm{d}x_i \, \frac{\eta}{\eta^2 - \prod_{j=1}^{d}(\cos[x_j/2])^2},$$

and another two similar integrals, where $\eta \geq 1$, are evaluated in an exact way in terms of generalized hypergeometric functions. A simple formula connecting $I_d$ and $J_{d+1}$ is given along with the differential equations for $I_d(\eta)$ and $J_d(\eta)$. An application of $I_d$ and $J_d$ in the theory of the Heisenberg antiferromagnet is discussed, together with possible generalizations to non-integer values of $d$. Corresponding integrals for sc lattices are also briefly reviewed.

# 1. Introduction

Watson-like integrals $I_2(\eta)$ and $I_3(\eta)$ appeared for the first time in the theory of magnetism in Kubo's classical paper [1], where the Heisenberg antiferromagnet (HAFM) was treated within linear spin wave theory (SWT). Kubo has shown that the spontaneous magnetisation of HAFM on the square and bcc lattices at $T = 0$ K is determined by these types of integrals. Later on, the development of the random phase approximation (RPA) in the Green's functions (GF) approach to HAFM, has led to the introduction of $J_d(\eta)$ in the expression for Néel temperature of HAFM on a $d$-dimensional bcc lattice (see [2, 3, 4, 5] and references therein).

The integrals $I_d(\eta)$ and $J_d(\eta)$ are analogous to the Watson integrals for ferromagnetic lattices. Definitions, exact solutions, applications and some approximate methods for calculating Watson integrals can be found in the original papers [6, 7, 8, 9, 10, 11, 12, 13, 14, 15] and the standard textbook [16].

In the last couple of decades there has been an increasing interest in antiferromagnets, mostly because of the discovery of the high temperature superconductivity in doped HAFMs. Various approximate methods and numerical simulations have been applied to these systems. A detailed analysis shows that at $T = 0$ K the SWT gives an excellent agreement with Monte Carlo simulations (MCS) [17]. Also, at $T = 0$ K, the GF approach within RPA, still using the integral $I_d(\eta)$, recovers SWT results in the first approximation. On the other hand, compared to the other GF approximations, at high temperatures the RPA gives best agreement with MCS [18] and experimental results [5].

Having all this in mind, and the fact that these integrals continue to appear in recent works [5, 19], it is of interest to have their exact solutions. It should be noted that $I_d(\eta)$ and $J_d(\eta)$ are not specific to either the GF or SWT approach. For example, they appear in the bosonic path integral theory of HAFM on $d$-dimensional bcc lattice (for the case of $d$-dimensional sc lattice see [20]).

In all previous papers, the integrals $I_d(\eta)$ and $J_d(\eta)$ have been treated in an approximate way (an exception is $I_2(1)$, calculated by Watson himself [6]). Although modern numerical methods allow one to calculate $I_d(\eta)$ and $J_d(\eta)$ with a high degree of precision, it is obvious that exact solution has its advantages. First, with the exact solution, no question arises as to the convergence of approximate expansions. Also, an exact solution often allows certain generalizations.

In the first three Sections, the integrals $I_d(\eta)$ and $J_d(\eta)$ are evaluated in an exact way, using the generalized hypergeometric function (GHGF) for an arbitrary integer $d$, giving at the same time the rigorous convergence conditions. As to our knowledge, these forms of $I_d(\eta)$ and $J_d(\eta)$ have never been reported before. Asymptotic behaviour for large $d$ and possible generalizations to continuous $d$ are investigated in Section 4. The application of $I_d(\eta)$ and $J_d(\eta)$ to HAFM is discussed in Section 5. Finally, in the Appendix, the method developed for evaluating $I_d(\eta)$ and $J_d(\eta)$ is used to calculate the lattice Green function for $d = 2$ and obtain a summation formula for the hypergeometric function.

## 2. Evaluation of $I_d(\eta)$

The $d$-dimensional Watson-like integral for bcc antiferromagnetic (AFM) lattice, $I_d(\eta)$, is defined as:

$$I_d(\eta) = \frac{1}{\pi^d} \prod_{i=1}^{d} \int_0^\pi \mathrm{d}x_i \; \frac{\eta}{\sqrt{\eta^2 - \prod_{j=1}^{d}(\cos[x_j/2])^2}}, \tag{1}$$

where $\eta \geq 1$ is a real parameter. Let us consider first $I_1(\eta)$. Being just the area under the graph of $[1 - 1/\eta^2 \cos^2 x]^{-1/2}$ for $0 \leq x \leq \pi/2$ and $\eta > 1$, this integral is equivalent to the usual definition of the complete elliptic integral of the first kind, i.e.

$$I_1(\eta) = \frac{2}{\pi} \int_0^{\pi/2} \mathrm{d}x \; \frac{1}{\sqrt{1 - \frac{1}{\eta^2}\cos^2 x}} = \frac{2}{\pi} \mathcal{K}\left(\frac{1}{\eta}\right). \tag{2}$$

A different form of $I_1(\eta)$ will be needed for further calculations. To this end, it is rewritten in terms of the hypergeometric function [21]

$$I_1(\eta) = {}_2F_1\left(\frac{1}{2}, \frac{1}{2}; 1; \frac{1}{\eta^2}\right). \tag{3}$$

For $\eta = 1$, the integral $I_1(\eta)$ does not exist.

The next step is to write $I_2(\eta)$ as

$$I_2(\eta) = \frac{4}{\pi^2} \int_0^{\pi/2} \mathrm{d}x \int_0^{\pi/2} \frac{\mathrm{d}y}{\sqrt{1 - a_x^2 \cos^2 y}}, \tag{4}$$

where

$$a_x = \frac{\cos x}{\eta}. \tag{5}$$

Using (3) and the fact that $a_x \leq 1$ for $\eta \geq 1$, we have

$$\begin{aligned} I_2(\eta) &= \frac{2}{\pi} \int_0^{\pi/2} \mathrm{d}x \; {}_2F_1\left(\frac{1}{2}, \frac{1}{2}; 1; a_x^2\right) \\ &= \frac{2}{\pi} \sum_{m=0}^{\infty} \frac{(1/2)_m (1/2)_m}{(1)_m} \frac{1}{m! \, \eta^{2m}} \int_0^{\pi/2} \mathrm{d}x \cos^{2m} x. \end{aligned} \tag{6}$$

In the last line, a series expansion for the hypergeometric function is used, and standard notation for the Pochhammer symbol is introduced:

$$(a)_m = \frac{\Gamma(a+m)}{\Gamma(a)}. \tag{7}$$

Since one can find in [22] the integral

$$\int_0^{\pi/2} \mathrm{d}z \; \cos^{2m} z = \frac{\sqrt{\pi}}{2} \frac{\Gamma(m+1/2)}{\Gamma(m+1)}, \tag{8}$$

and

$$\frac{\Gamma(m+1/2)}{\Gamma(m+1)} = \frac{(1/2)_m}{(1)_m}\sqrt{\pi}, \tag{9}$$

the expression for $I_2(\eta)$ reduces to

$$I_2(\eta) = \sum_{m=0}^{\infty} \frac{(1/2)_m \, (1/2)_m \, (1/2)_m}{(1)_m \, (1)_m} \, \frac{1}{m! \, \eta^{2m}}. \tag{10}$$

For $\eta \geq 1$, the series from (10) converges absolutely and defines the GHGF [23]:

$$I_2(\eta) = {}_3F_2\left(\frac{1}{2}, \frac{1}{2}, \frac{1}{2}; 1, 1; \frac{1}{\eta^2}\right). \tag{11}$$

The procedure for evaluating $I_3(\eta)$ is almost the same. One can write $I_3(\eta)$ as

$$I_3(\eta) = \frac{2}{\pi} \int_0^{\pi/2} dz \, \frac{4}{\pi^2} \int_0^{\pi/2} dx \int_0^{\pi/2} \frac{dy}{\sqrt{1 - b_{xz}^2 \cos^2 y}},$$

$$b_{xz} = \frac{\cos x \, \cos z}{\eta}. \tag{12}$$

and repeat all the steps that led to (11), to obtain

$$I_3(\eta) = {}_4F_3\left(\frac{1}{2}, \frac{1}{2}, \frac{1}{2}, \frac{1}{2}; 1, 1, 1; \frac{1}{\eta^2}\right). \tag{13}$$

where ${}_4F_3$ denotes a GHGF. As can be seen from (3), (11) and (13), the general solution of $I_d(\eta)$ is

$$I_d(\eta) = \frac{1}{\pi^d} \prod_{i=1}^d \int_0^\pi dx_i \, \frac{\eta}{\sqrt{\eta^2 - \prod_{j=1}^d (\cos[x_j/2])^2}}$$

$$= {}_{d+1}F_d\left(\underbrace{\frac{1}{2}, \frac{1}{2}, ..., \frac{1}{2}}_{d+1 \text{ times}}; \underbrace{1, 1, ..., 1}_{d \text{ times}}; \frac{1}{\eta^2}\right), \qquad \eta \geq 1, \, d > 1, \tag{14}$$

and it can be easily proved by induction. It is important to notice that if $p = q + 1$, the series defining GHGF

$$_pF_q\left(\underbrace{a_1, a_2, \cdots, a_p}_{p \text{ times}}; \underbrace{b_1, b_2, \cdots, b_q}_{q \text{ times}}; \frac{1}{\eta^2}\right) = \sum_{m=0}^{\infty} \frac{(a_1)_m \, (a_2)_m \cdots (a_p)_m}{(b_1)_m \, (b_2)_m \cdots (b_q)_m} \, \frac{1}{m! \, \eta^{2m}} \tag{15}$$

converges absolutely only for $\eta > 1$. If $\sum_j b_j > \sum_i a_i$, the series converges also for $\eta = 1$. The required conditions for $b_j$'s are that $b_j \neq 0$ and $b_j \notin \mathbb{Z}_-$ [23].

Another similar $d$-dimensional integral is

$$\widetilde{I_d}(\eta) = \frac{1}{\pi^d} \prod_{i=1}^d \int_0^\pi dx_i \, \sqrt{\eta^2 - \prod_{j=1}^d (\cos[x_j/2])^2}$$

$$= \eta \, {}_{d+1}F_d\left(-\frac{1}{2}, \underbrace{\frac{1}{2}, \frac{1}{2}, ..., \frac{1}{2}}_{d \text{ times}}; \underbrace{1, 1, ..., 1}_{d \text{ times}}; \frac{1}{\eta^2}\right). \tag{16}$$

To derive (16) one needs to use the complete elliptic integral of the second kind and its hypergeometric function representation [21]

$$\mathcal{E}(t) = \frac{\pi}{2} F\left(-\frac{1}{2}, \frac{1}{2}; 1; t^2\right), \tag{17}$$

repeating the outlined procedure. Since $I_d(\eta) = \partial \widetilde{I}_d(\eta)/\partial \eta$, the integration proves (16).

As a second example we present solution of the $d$-dimensional integral $\widetilde{J}_d(\eta)$

$$\widetilde{J}_d(\eta) = \frac{1}{\pi^d} \prod_{i=1}^{d} \int_0^\pi dx_i \, \frac{\prod_{j=1}^{d}(\cos[x_j/2])^2}{\sqrt{\eta^2 - \prod_{j=1}^{d}(\cos[x_j/2])^2}}$$

$$= \eta \,_{d+1}F_d\left(\underbrace{\frac{1}{2},...,\frac{1}{2}}_{d+1 \text{ times}}; \underbrace{1,...,1}_{d \text{ times}}; \frac{1}{\eta^2}\right) - \eta \,_{d+1}F_d\left(-\frac{1}{2}, \underbrace{\frac{1}{2}...,\frac{1}{2}}_{d \text{ times}}; \underbrace{1,...,1}_{d \text{ times}}; \frac{1}{\eta^2}\right)$$

$$= \frac{1}{\eta}\left(\frac{1}{2}\right)^d \,_{d+1}F_d\left(\frac{1}{2}, \underbrace{\frac{3}{2},\frac{3}{2}...,\frac{3}{2}}_{d \text{ times}}; \underbrace{2,2,...,2}_{d \text{ times}}; \frac{1}{\eta^2}\right). \tag{18}$$

In deriving (18), a series expansion for the GHGF is used, together with the fact that $(a)_{m+1} = (a+1)_m \Gamma(a+1)/\Gamma(a)$ and $\Gamma(3/2)/\Gamma(1/2) = 1/2$. The integrals (16) and (18) appear in the GF theory of HAFM on hyper bcc lattice within Callen's approximation [24]. In the sequel we will focus on the RPA GF.

The integrals similar to the isotropic case $\eta = 1$ of $\widetilde{J}_d(\eta)$, but with more input parameters, were considered in [25] from general and formal mathematical points of view. It should be noted that these integrals are reduced to the very well-poised hypergeometric function (Zudilin's theorem), demanding that $a_2 = a_1/2 + 1$ for the coefficients in (15), which is not the case for any of the integrals from the present paper (for a precise definition of the very well-poised hypergeometric function and different proof of Zudilin's theorem, see [26]). Another application of GHGF to the calculation of integrals in physical problems can be found in [27], where the authors discussed a certain class of 1D integrals that appear in classical mechanics and WKB approximation of quantum mechanics.

Since the integrals $I_d(\eta)$ are expressed in terms of GHGF, it is found that they are solutions of the differential equation

$$\left[\widehat{\theta}^{d+1} - \frac{1}{\eta^2}(\widehat{\theta}-1)^{d+1}\right] I_d(\eta) = 0, \tag{19}$$

where $\widehat{\theta} = \eta \partial_\eta$.

## 3. Evaluation of $J_d(\eta)$

The $d$-dimensional AFM bcc lattice integral $J_d(\eta)$ is defined as

$$J_d(\eta) = \frac{1}{\pi^d} \prod_{i=1}^{d} \int_0^{\pi} dx_i \, \frac{\eta}{\eta^2 - \prod_{j=1}^{d}(\cos[x_j/2])^2}. \tag{20}$$

Just as in the previous section, the $d = 1$ case is analysed first. It is useful to introduce the following indefinite integral:

$$\tilde{J}(y, a) = \int \frac{dy}{1 - a^2 \cos^2 y} = \frac{2}{1 - a^2} \int dt \, \frac{1 + t^2}{\prod_{\alpha=1}^{4}(t - t_\alpha)}, \tag{21}$$

This integral can be found in standard tables (e.g [22]), but for completeness of the paper, we sketch here its solution. In (21) $y = \tan[t/2]$, $a < 1$ and $t_\alpha$ are the roots of the equation:

$$t^4(1 - a^2) + t^2 2(1 + a^2) + (1 - a^2) = 0. \tag{22}$$

Being biquadratic, (22), it is easily solved:

$$t_1 = i\sqrt{\frac{1 + a}{1 - a}}, \quad t_2 = -t_1, \quad t_3 = -\frac{1}{t_1}, \quad t_4 = \frac{1}{t_1}. \tag{23}$$

Using factorisation, elementary integration and a few of trigonometric identities, one obtains $\tilde{J}(y, a)$ as given in [22]

$$\tilde{J}(y, a) = \frac{1}{\sqrt{1 - a^2}} \arctan \frac{\tan[y/2]}{\sqrt{1 - a^2}}. \tag{24}$$

Putting $\eta = 1/a > 1$ and taking appropriate limits, one finds that

$$J_1(\eta) = \frac{1}{\eta \sqrt{1 - (1/\eta)^2}}. \tag{25}$$

For $d = 2$, the integral $J_2(\eta)$ can be written as

$$J_2(\eta) = \frac{4}{\pi^2 \eta} \int_0^{\pi/2} dx \int_0^{\pi/2} \frac{dy}{1 - a_x^2 \cos^2 y}, \tag{26}$$

with $a_x$ defined in (5). Now, using (25), (3) and (2) one finds

$$J_2(\eta) = \frac{1}{\eta} \, _2F_1\left(\frac{1}{2}, \frac{1}{2}; 1; \frac{1}{\eta^2}\right). \tag{27}$$

Evaluation of $J_3$ is straightforward

$$J_3(\eta) = \frac{2}{\pi \eta} \int_0^{\pi/2} dz \, \frac{4}{\pi^2} \int_0^{\pi/2} dx \int_0^{\pi/2} \frac{dy}{1 - b_{xz}^2 \cos^2 y}$$

$$= \frac{1}{\eta} \, _3F_2\left(\frac{1}{2}, \frac{1}{2}, \frac{1}{2}; 1, 1; \frac{1}{\eta^2}\right), \tag{28}$$

where $b_{xz}$ is defined in (12). The solution for the general integer $d$ now reads

$$J_d(\eta) = \frac{1}{\pi^d} \prod_{i=1}^{d} \int_0^{\pi} dx_i \, \frac{\eta}{\eta^2 - \prod_{j=1}^{d}(\cos[x_j/2])^2}$$

$$= \frac{1}{\eta} \, {}_dF_{d-1}\left(\underbrace{\frac{1}{2}, \frac{1}{2}, ..., \frac{1}{2}}_{d \text{ times}}; \underbrace{1, 1, ..., 1}_{d-1 \text{ times}}; \frac{1}{\eta^2}\right), \quad (29)$$

and this can be proved by induction. The convergence criteria for $J_d(\eta)$ are the same as those for $I_d(\eta)$. Furthermore, by comparing (14) and (29) one finds a simple connection between $J_d(\eta)$ and $I_d(\eta)$:

$$I_d(\eta) = \eta \, J_{d+1}(\eta), \quad (30)$$

or, equivalently

$$J_d(\eta) = \frac{1}{\eta} I_{d-1}(\eta). \quad (31)$$

It should be pointed out that, to our knowledge, the forms (14), (16), (18) and (29), together with (30) or (31), have never been reported before.

The differential equation for $J_d(\eta)$ has a somewhat more complicated general form. From (31) and (19) it follows that $\eta J_d(\eta)$ satisfies

$$\left[\widehat{\theta}^d - \frac{1}{\eta^2}(\widehat{\theta} - 1)^d\right] [\eta J_d(\eta)] = 0. \quad (32)$$

Now, since

$$\widehat{\theta}^d f(\eta) = \sum_{\alpha=0}^{d} S(d, \alpha) \eta^\alpha \partial_\eta^\alpha f(\eta), \quad (33)$$

where $S(d, \alpha)$ are the Stirling numbers of the second kind and

$$(\widehat{\theta} - 1)^d f(\eta) = \sum_{k=0}^{d} \binom{d}{k} (-1)^{d-k} \widehat{\theta}^k f(\eta), \quad (34)$$

with the binomial coefficients $\binom{d}{k}$, one can find the differential equation for $J_d(\eta)$:

$$\left[\eta \widehat{\theta}^d + \sum_{\alpha=1}^{d} S(d, \alpha) \, \alpha \, \eta^\alpha \partial_\eta^{\alpha-1} - \frac{1}{\eta}(\widehat{\theta} - 1)^d \right.$$
$$\left. - \frac{1}{\eta^2} \sum_{k=0}^{d} \sum_{\alpha=1}^{k} \binom{d}{k} (-1)^{d-k} S(k, \alpha) \, \alpha \, \eta^\alpha \partial_\eta^{\alpha-1}\right] J_d(\eta) = 0. \quad (35)$$

This equation is valid for an arbitrary $d$. For $d = 3$, it reduces to

$$\left[\eta^2(\eta^2 - 1)\partial_\eta^3 + 3\eta(2\eta^2 - 1)\partial_\eta^2 + (7\eta^2 - 1)\partial_\eta + \eta\right] J_3(\eta) = 0, \quad (36)$$

which is just the equation for the cubic lattice Green function

$$G^{(2)}(\eta) = \frac{1}{\eta} \, {}_3F_2\left(\frac{1}{2}, \frac{1}{2}, \frac{1}{2}; 1, 1; \frac{1}{\eta^2}\right),$$

as it is given in [10].

## 4. Asymptotic behaviour for large $d$ and further generalizations

To investigate the asymptotic behaviour for $d \gg 1$, it is convenient to rewrite (14) as

$$I_d(\eta) = \sum_{m=0}^{\infty} \frac{[(1/2)_m]^{d+1}}{[(1)_m]^d} \frac{1}{m!\,\eta^{2m}} = \left(\frac{1}{\sqrt{\pi}}\right)^{d+1} \sum_{m=0}^{\infty} \frac{[\Gamma(m+1/2)]^{d+1}}{[\Gamma(m+1)]^d} \frac{1}{m!\,\eta^{2m}}. \tag{37}$$

Since $d \gg 1$, the main contribution to $I_d$ comes from the lowest terms in the above expansion. Thus

$$I_d(\eta) = 1 + \frac{1}{\eta^2}\left(\frac{1}{2}\right)^{d+1} + \frac{1}{\eta^4}\left(\frac{3}{8}\right)^{d+1} + \frac{1}{\eta^6}\left(\frac{5}{16}\right)^{d+1} + \frac{1}{\eta^8}\left(\frac{35}{128}\right)^{d+1} +$$

$$+ \frac{1}{\eta^{10}}\left(\frac{63}{256}\right)^{d+1} + \frac{1}{\eta^{12}}\left(\frac{231}{1024}\right)^{d+1} + \frac{1}{\eta^{14}}\left(\frac{429}{2048}\right)^{d+1} + \cdots \tag{38}$$

for $d \gg 1$. This means that

$$\lim_{d \to \infty} I_d(\eta) = 1, \tag{39}$$

for any $\eta \geq 1$.

The series from (37) is well defined, even for a non-integer $d$, as we shall see below. For practical computations, one can define $I_d(\eta)$ depending continuously on the parameter $d$ as a power series in $1/\eta^2$

$$I(d, \eta) = \sum_{m=0}^{\infty} \beta(d, m) \left(\frac{1}{\eta^2}\right)^m, \quad \beta(d, m) = \frac{[\Gamma(m+1/2)/\sqrt{\pi}]^{d+1}}{m!\,[\Gamma(m+1)]^d} = \left[\frac{(1/2)_m}{(1)_m}\right]^{d+1}. \tag{40}$$

The notation $I(d, \eta)$ indicates that $d$ can assume non-integer values. Consider the case $\eta = 1$ first. Convergence of $I(d, 1)$ can be confirmed using Raabe's test [21], which states that the series $\sum_{n=1}^{\infty} a_n$ converges if for a large nough $n$

$$n\left(\frac{a_n}{a_{n+1}} - 1\right) > 1. \tag{41}$$

Since $\beta(d, n+1) = (n+1/2)^{d+1}\,(n+1)^{-d-1}\beta(d, n)$, the convergence condition for $I(d, 1)$ is obtained from (41)

$$\frac{d+1}{2} > 1. \tag{42}$$

In other words, the series

$$I(d, 1) = \sum_{m=0}^{\infty} \beta(d, m) \tag{43}$$

converges for any positive real $d > 1$. For $d = 1$ and $\eta > 1$ the series (40) reduces to the hypergeometric function ${}_2F_1(1/2, 1/2; 1; 1/\eta^2) = 2\mathcal{K}(1/\eta^2)/\pi$. Since for any $d > 1$

$$\beta(d, m) \leq \frac{(1/2)_m\,(1/2)_m}{(1)_m\,m!}, \tag{44}$$

by the Weierstrass criterion [21] we conclude that the series (40) converges for any $d \geq 1$ and $\eta > 1$. Moreover, it can be shown that for $\eta = 1$ and $d > 1$ or $\eta > 1$ and $d \geq 1$, the series

$$I'(d, \eta) = \sum_{m=0}^{\infty} \beta(d, m)\,\ln\left[\frac{\Gamma(m+1/2)}{\Gamma(m+1)\sqrt{\pi}}\right]\left(\frac{1}{\eta^2}\right)^m \tag{45}$$

converges uniformly. Thus, $I(d,\eta)$ can be considered indeed as a smooth function of the continuous parameter $d$. For the integer values of $d$, it reduces to the GHGF $I_d(\eta)$ given in (14).

A plot of $I(d,\eta)$, continuously depending on $d$, is given in Figure 1. For concreteness, we take $\eta = 1$ and $\eta = 1.005$ (see section 5). In calculating $I(d,\eta)$, an approximate formula

$$I(d,\eta) = \sum_{m=0}^{M} \beta(d,m) \left(\frac{1}{\eta^2}\right)^m, \qquad (46)$$

is used, with $M = 10^4$. For $\eta = 1$, the series (40) converges slowly and near the divergent point ($d = \eta = 1$) the last term included contributes by $3.02 \cdot 10^{-5}$. As $d$ grows, the convergence is getting faster and the truncation error reduces. When $\eta > 1$, the convergence is much faster and truncation error is $< 1.42 \cdot 10^{-48}$. Obviously, to investigate the singular behaviour of $I(d,\eta)$, more sophisticated methods for summing (40) must be used.

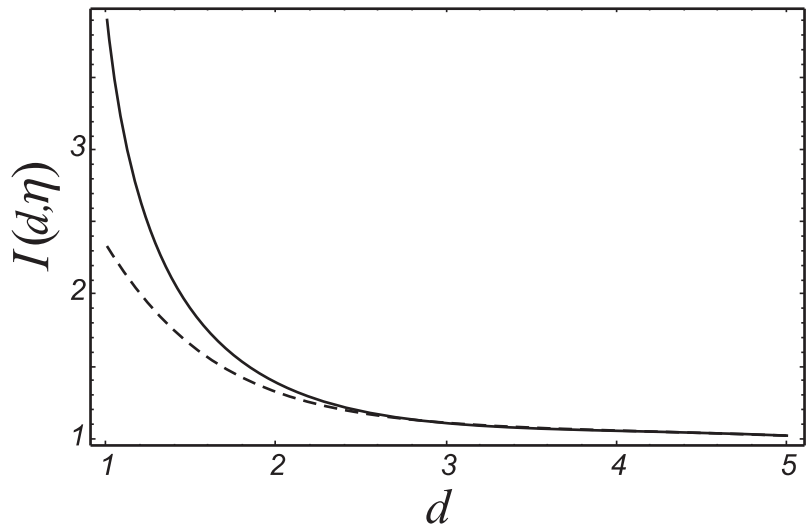

Figure 1: Watson-like integrals $I(d,\eta)$ (full line: $\eta = 1$; dashed line: $\eta = 1.005$) as functions of a continuous variable $d$ (dimensionality of the lattice) calculated by (46).

As with the other extrapolations to a non-integer $d$ [28, 29], this choice may look natural, but different kinds of generalizations may be possible.

In the next section we apply the formula (40) to HAFM.

## 5. Application to the Heisenberg antiferromagnet

In the theory of HAFM, the parameter $\eta$ represents spin anisotropy, defining the easy axis of magnetisation [5, 30]. The case $\eta = 1$ corresponds to an isotropic HAFM. The particular case of a rather small anisotropy $\eta = 1.005$, dealt with in our previous study [5], is also considered.

The expressions for ground state spontaneous magnetisation and critical temperature of 2D HAFM with spin $XXZ$ anisotropy are derived in our previous paper [5] by means of RPA GF. In that paper, the integral $I_2(\eta)$ was denoted by $2P_S(0)+1$ and $J_2(\eta)$ by $C_2$. The results from [5] can be readily extended to an arbitrary $d$-dimensional bcc lattice by replacing $I_2(\eta)$ and $J_2(\eta)$ with $I_d(\eta)$ and $J_d(\eta)$. Using (19), (20), (23) and (24) from [5], one obtains the ground-state magnetisation

$$\langle \hat{S} \rangle_0^d(\eta) = \frac{[S - P_S^d(\eta)][1 + P_S^d(\eta)]^{2S+1} + [S + 1 + P_S^d(\eta)][P_S^d(\eta)]^{2S+1}}{[1 + P_S^d(\eta)]^{2S+1} - [P_S^d(\eta)]^{2S+1}},$$

$$P_S^d(\eta) = \frac{1}{2}[I_d(\eta) - 1], \qquad (47)$$

and critical temperature for arbitrary spin $S$ of the hyper bcc AFM lattice:

$$k_{\rm B} T_{\rm N}^d(\eta) = \frac{S(S+1)}{3} \frac{z(d) \, J}{J_d(\eta)}. \qquad (48)$$

Here, $J$ denotes the nearest neighbour exchange integral and $z(d) = 2^d$ is the number of nearest neighbours.

Consider $d = 1$ and $\eta = 1$ first. The $I_1(1)$ diverges and the ground state of isotropic linear chain becomes disordered ($\langle \hat{S} \rangle_0^1(1) \to 0$). Since in the case of $d = 2$ and $\eta = 1$, $I_2(1) = 1.3932$, due to the quantum fluctuations, the 2D system exhibits long-range order at $T = 0$K with $\langle \hat{S} \rangle_0^2(1) < S$. However, $J_2(1)$ diverges, and for the isotropic 2D system there can be no long-range order at finite temperatures. This is a rigorous proof that in this case the RPA GF solution obeys the Mermin-Wagner theorem [32]. When $\eta \neq 1$, the long-range order in 2D HAFM is destroyed at a finite temperature determined by (48), for $d = 2$ (more details on the role of spin anisotropy are given in [5]). As expected, for $d > 2$, the state of isotropic and anisotropic bcc HAFM's displays the long-range order for $T \neq 0$, up to $T_{\rm N}$.

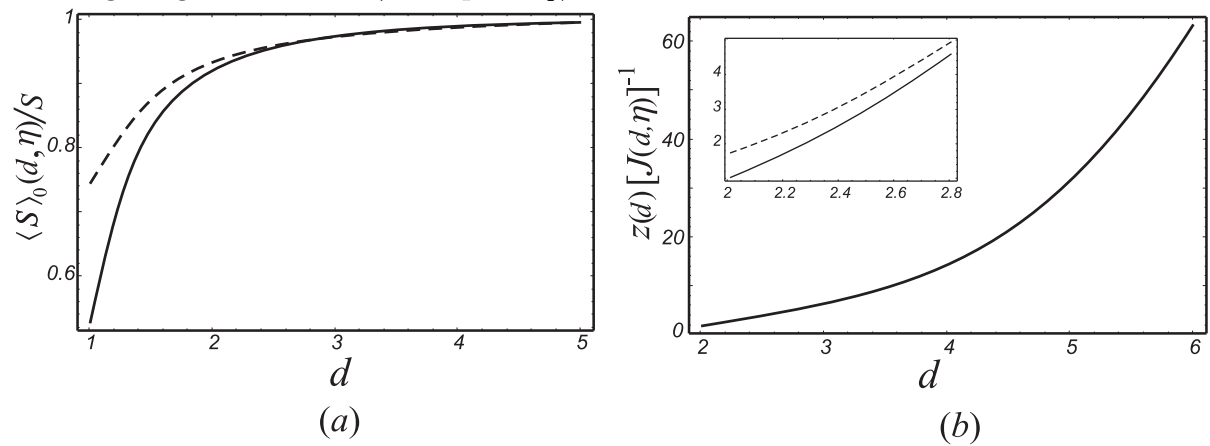

(a)  (b)

Figure 2: (a) Relative spontaneous magnetisation $\langle \hat{S} \rangle_0(d,\eta)/S$, and (b) reduced critical temperature $3T_{\rm N}(d,\eta)[JS(S+1)]^{-1} = z(d)[J(d,\eta)]^{-1}$ of the HAFM on hyper bcc lattice as a function of the continuous variable $d$ (dimensionality of lattice). Inset on the graph (b) shows the region near $d = 2$. Full lines correspond to $\eta = 1$, and dashed lines to $\eta = 1.005$. Spin $S = 5/2$.

Following the ideas of quantum field theory and renormalization group [28, 31], let us suppose that thermodynamic parameters of the bcc HAFM depend continuously on the dimension $d$. To generalize previous discussion to a non-integer $d$, use was made of (40). By replacing $I_d(\eta)$ with $I(d, \eta)$ in (47) and (48) one obtains the ground-state spontaneous magnetisation $\langle \hat{S} \rangle_0(d, \eta)$ and the critical temperature $k_B T_N(d, \eta)$ of the HAFM on hyper bcc lattice with $d$ as a continuous parameter. The function $J_d(\eta)$ was extended to $J(d, \eta) = I(d-1, \eta)\eta^{-1}$, according to (31).

The results obtained in this manner for $S = 5/2$, $\eta = 1$ and $\eta = 1.005$ are shown in Figure 2.

It follows from (39), (47) and (48) that $\langle \hat{S} \rangle_0(d, \eta) \to S$ as $d \to \infty$. This means that quantum fluctuations are suppressed in the limit of an infinite number of space dimensions. On the other hand, in the same limit $k_B T_N(d, \eta) \sim 2^d$, and one can see that for an infinite-dimensional hyper bcc lattice HAFM the classical phase transition is absent.

At this point, it is not clear whether this kind of extrapolation (using generalized hypergeometric function) to a non-integer $d$ can be applied to the Heisenberg antiferromagnet and ferromagnet model on an arbitrary lattice, but some examples for sc lattices are discussed in the Appendix. However, by extending the results from [10, 11, 12] to arbitrary integer $d$, one obtains

$$G_d^{(2)}(\eta) = \frac{1}{\pi^d} \prod_{i=1}^{d} \int_0^\pi \mathrm{d}x_i \, \frac{1}{\eta - \prod_{j=1}^{d} \cos x_j} = \frac{1}{\eta} \,_dF_{d-1}\left(\underbrace{\frac{1}{2}, \frac{1}{2}, ..., \frac{1}{2}}_{d \text{ times}}; \underbrace{1, 1, ..., 1}_{d-1 \text{ times}}; \frac{1}{\eta^2}\right) \quad (49)$$

and, by similar arguments, can find the RPA Curie temperature for the Heisenberg ferromagnet on a $d$-dimensional bcc lattice [2, 4, 7]:

$$k_B T_C^d(\eta) = \frac{S(S+1)}{3} \frac{z(d) \, J}{G_d^{(2)}(\eta)} = k_B T_N^d(\eta), \quad (50)$$

since $G_d^{(2)}(\eta) = J_d(\eta)$. It is obvious from (50) that all comments made for the Néel temperature also apply for the Curie temperature of a Heisenberg ferromagnet on a $d$-dimensional bcc lattice.

**Appendix**

Here we discuss some integrals for $d$-dimensional sc lattices. The method developed in Sections 2 and 3 does not give solutions in closed form for sc lattices. One can still apply this method to obtain the relation between hypergeometric functions. Consider the cubic lattice Green function [9, 10]

$$G_2^{(1)}(\eta) = \frac{1}{\pi^2} \int_0^\pi \mathrm{d}x \int_0^\pi \mathrm{d}y \frac{1}{\eta - (\cos x + \cos y)/2}, \quad (A.1)$$

which is just a Watson integral for sc lattice and $d = 2$. This integral can also be written as

$$G_2^{(1)}(\eta) = \frac{2}{\pi^2} \int_0^\pi \mathrm{d}x \, c_x \int_0^\pi \frac{\mathrm{d}y}{1 - c_x \cos y} = \frac{2}{\pi} \int_0^\pi \mathrm{d}x \frac{c_x}{\sqrt{1 - c_x^2}},$$
$$c_x = (2\eta - \cos x)^{-1}. \tag{A.2}$$

Using

$$\frac{1}{(1-x)^k} = {}_2F_1(k, \, 1 \,; 1; \, x), \tag{A.3}$$

together with $(a)_{2n} = 2^{2n}(a/2)_n(a/2 + 1/2)_n$ and (8), one can show that

$$\int_0^\pi \frac{\mathrm{d}x}{(1 - \eta^{-1} \cos x)^k} = \pi \, {}_2F_1\left(\frac{k+1}{2}, \, \frac{k}{2} \,; 1; \, \frac{1}{\eta^2}\right), \tag{A.4}$$

and the solution for $G_2^{(1)}(\eta)$ can be expressed in terms of the infinite series

$$G_2^{(1)}(\eta) = \frac{1}{\eta} \sum_{n=0}^\infty \frac{(1/2)_n}{n! \, (4\eta^2)^n} \, {}_2F_1\left(n + \frac{1}{2}, \, n + 1 \,; 1; \, \frac{1}{4\eta^2}\right). \tag{A.5}$$

Since the bcc and sc lattices are equivalent for $d = 2$, comparing (49) and (A.5) one finds the following summation formula for the hypergeometric function

$$\sum_{n=0}^\infty \frac{(1/2)_n}{n! \, (4\eta^2)^n} \, {}_2F_1\left(n + \frac{1}{2}, \, n + 1 \,; 1; \, \frac{1}{4\eta^2}\right) = {}_2F_1\left(\frac{1}{2}, \, \frac{1}{2} \,; 1; \, \frac{1}{\eta^2}\right). \tag{A.6}$$

We were not able to obtain similar expressions for $d > 2$.

As far as the extrapolating to a non-integer $d$ for sc integrals is concerned, one can use modifications of the procedure proposed by Maradudin [33, 9] for ferromagnetic lattices:

$$\frac{1}{\pi^d} \prod_{i=1}^d \int_0^\pi \mathrm{d}x_i \, \frac{1}{\sqrt{\eta^2 - (d^{-1} \sum_{j=1}^d \cos x_j)^2}} = \frac{d}{\pi} \int_0^\infty \int_0^\infty \mathrm{d}t \, \mathrm{d}s \, \frac{\mathrm{e}^{-\eta(s+t)d}}{\sqrt{ts}} \, [I_0(t - s)]^d, \tag{A.7}$$

$$\frac{1}{\pi^d} \prod_{i=1}^d \int_0^\pi \mathrm{d}x_i \, \frac{1}{\eta^2 - (d^{-1} \sum_{j=1}^d \cos x_j)^2} = d^2 \int_0^\infty \int_0^\infty \mathrm{d}t \, \mathrm{d}s \, \mathrm{e}^{-\eta(s+t)d} \, [I_0(t - s)]^d, \tag{A.8}$$

where $I_0(x)$ is the modified Bessel function of the first kind.

### Acknowledgments

This work was supported by the Ministry of Science and Technological Development of the Republic of Serbia (Grant No 141018). The authors also wish to thank the referees for their helpful comments.